\documentclass[useAMS,usenatbib]{mn2e}

\usepackage{aas_macros}
\usepackage{graphicx}
\usepackage{color}
\usepackage{multirow}

\newcommand{\der}[2]{\frac{\mathrm{d#1}}{\mathrm{d#2}}}
\newcommand{\e}[1]{\times 10^{#1}}
\renewcommand{\l}{\ell}
\newcommand{\ellp}{{\ell^\prime}}
\newcommand{\healpix}{HEALPix }
\newcommand{\Planck}{\textit{Planck} }

\title[A new model for NVSS]{A new model for NVSS galaxy catalogue using the redshift distribution and the halo minimum mass}
\author[A. Marcos-Caballero et al.]{A. Marcos-Caballero$^{1,2}$\thanks{E-mail:
marcos@ifca.unican.es}, P. Vielva$^{1}$,
  E. Mart\'inez-Gonz\'alez$^{1}$, F. Finelli$^{3,4}$, \newauthor
  A. Gruppuso$^{3,4}$ and F. Schiavon$^{3,4}$ \\
$^{1}$Instituto de F\'isica de Cantabria (CSIC - Univ. Cantabria),
  Av. de Los Castros s/n, E-39005 - Santander, Spain \\ 
$^{2}$Departamento de F\'isica Moderna (Univ. Cantabria), Av. de Los
Castros s/n, E-39005 - Santander, Spain \\
$^{3}$INAF-IASF Bologna, Istituto di Astrofisica Spaziale e Fisica Cosmica di Bologna, Via Gobetti 101, I-40129 Bologna, Italy \\
$^{4}$INFN, Sezione di Bologna, Via Berti Pichat 6/2, I-40127 Bologna, Italy}
\begin{document}

\date{Accepted ???. Received ???; in original form \today}

\pagerange{\pageref{firstpage}--\pageref{lastpage}} \pubyear{2002}

\maketitle

\label{firstpage}

\begin{abstract}

In the present paper we study the radio sources in the NRAO VLA SKY
Survey (NVSS) analysing its power spectrum and galaxy
distribution. There is a discrepancy between the theoretical models in
the literature and the data at large scales. A new model for NVSS is
proposed combining the power spectrum data from NVSS and the galaxy
distribution from the Combined EIS-NVSS Survey Of Radio Sources
(CENSORS). Taken into account these two data sets the differences in
the power spectrum at large scales are reduced, but there is still
some tension between the two data sets. Different models are compared
using Bayesian evidence. A model for the galaxy distribution based on
a gamma function provides a higher evidence against other models
proposed in the literature. In addition, the effect of primordial
non-Gaussianity has been also considered. The $2$-$\sigma$ constraints
on the local non-Gaussian $f_{\rmn{NL}}$ parameter are $-43 <
f_{\rmn{NL}} < 142$.

\end{abstract}

\begin{keywords}
-- cosmology
\end{keywords}

\section{Introduction}
The large-scale structure (LSS) of the universe is one of the most
important tools in modern cosmology, in particular, to probe the
physics of the very early universe. Galaxy surveys help us to trace
current matter density perturbations, which are directly related to
the initial curvature perturbations.  The statistical properties of
these perturbations (e.g., the power spectrum and the probability
distribution) are directly determined by the specific inflationary
model responsible for the exponential growth of the initial
perturbations~\citep[see, e.g.,][]{liddle2000}.

The standard paradigm of cosmic inflation predicts that the initial
perturbations are near scale-invariant and Gaussianly
distributed. This simplest model is confirmed by the current
observations of the cosmic microwave
background~\citep[CMB,][]{planck22_2013,planck24_2013}. Nevertheless,
some non-standard inflation scenarios are not ruled-out by current CMB
measurements. It is expected that the next step-forward on
constraining cosmic inflation is coming from future galaxy surveys as
Euclid~\citep{laureijs2011}, J-PAS~\citep{benitez2009},
LSST~\citep{lsst2009}, WFIRST~\citep{green2012} or
Big-Boss~\citep{schlegel2009}, both in terms of the power spectrum and
the probability distribution of the initial perturbations. Let us
remark that, despite these capabilities to constrain cosmic inflation,
the major scientific goal of future surveys is focused on the study of
dark energy, by exploiting the physics of the baryon acoustic
oscillations~\citep[see,][as an example for Euclid
forecasts]{amendola2013}.

This paper studies one of the most widely galaxy survey used up to
date: the NRAO VLA Sky Survey (NVSS)~\citep{condon1998}. The strength
of the NVSS radio galaxy catalogue comes from its large sky coverage
and its relatively high redshift range. As a drawback, it does not
posses individual redshift estimation of the sources, but just a
statistical description. These characteristics make NVSS a suitable
catalogue (from the cosmological point of view) for cross-correlation
studies with the CMB, and, also, to constrain the probability
distribution of the initial perturbations.

Regarding the former analyses, it is worth mentioning that NVSS was
cross-correlated with WMAP~\citep{larson2011} to report the first
detection of the integrated Sachs-Wolfe (ISW)
effect~\cite{boughn2004}. Many subsequent
analyses~\citep[e.g.,][]{vielva2006,pietrobon2006,mcewen2007,schiavon2012,giannantonio2012,barreiro2013}
further explored different aspects of this cross-correlation and,
recently, it was also used to study the ISW effect
on~\Planck~\citep{planck01_2013} data~\citep{planck19_2013}. The ISW
signal depends very much on the accurate description of the redshift
distribution of the galaxies. In addition, the actual sensitivity to
detect the ISW signal is also very much dependent on the correct
modelling of the NVSS angular power spectrum (which also depends on
the galaxy redshift distribution itself).

Constraining non-Gaussianity with galaxy surveys is being an active
topic since the last five
years~\citep[e.g.,][]{dalal2008,matarrese2008,desjacques2010,xia2011,giannantonio2013}. This
is a field in which NVSS has been also commonly used. Primordial
non-Gaussianity introduces a modification in the two-point correlation
function (or, conversely, the angular power spectrum) of dark matter
halos \citep{dalal2008,matarrese2008,desjacques2010}. An evidence of
this fact appears in the bias relation between halos and the
underlying matter. For instance, the linear deterministic bias $b$
becomes scale-dependent, and this dependence term is proportional to
the amount of primordial non-Gaussianity.

From the previous discussion, it is clear that understanding the
angular power spectrum and, therefore, the galaxy redshift
distribution, is a necessary condition to extract useful cosmological
information from NVSS. However, NVSS is known to suffer from some
systematics, especially at very large scales. These systematics are
mostly reflected as an excess of power at scales greater than $20$
degrees~\citep[e.g.,][]{hernandez2010}. The aim of this work is,
precisely, to study in detail the statistical properties of NVSS,
trying to explore further these incompatibilities.

The paper is organized as follows. In section 2 the theoretical
framework used in this work is discussed, introducing some basic
concepts. The data and their analysis are described in section 3 and
4. The results and conclusions are presented in sections 5 and 6
respectively.

\section{Theoretical framework}
According to the halo model, galaxies are formed inside halos and then
it is important to characterize the clustering of halos in order to
study the galaxy power spectrum. In next subsections it will be
discussed the mass function of halos, the bias and
redshift distribution of galaxies and the angular power spectum. In
the last subsection it will be introduced the effect of primordial
non-Gaussianity in the LSS.

\subsection{Mass function}

The mass function of halos gives the number of dark matter halos with
mass $m$ per unit of volume at redshift $z$. Theoretical predictions
of the mass function can be calculated assuming Gaussian perturbations
\citep{press1974}. It is useful to introduce the $\nu$ parameter
representing the peak height. It depends on the redshift and the mass
through the formula $\nu = \delta_c(z)/\sigma_m$, where $\sigma_m$ is
the variance of the linear matter perturbation smoothed at the scale
given by the mass $m$ (it is assumed a spherical top-hat filter) and
$\delta_c(z)$ is the critical overdensity for the collapse. The
Gaussian mass function $n_G(m,z)$ is written in the universal form
\citep{press1974}:
\begin{equation}
n_{G}(m,z) = \frac{\bar{\rho}}{m} f(\nu) \der{\nu}{m} \ .
\end{equation}
In this expression $\bar{\rho}$ is the mean matter density of the
universe and $f(\nu)$  is the multiplicity function. We adopt the
\citet{sheth1999} mass function where
\begin{equation}
f(\nu) = A \left( 1 + \frac{1}{(2a\nu^2)^p} \right)
\sqrt{\frac{a}{2\pi}} e^{-a\nu^2/2} \ .
\end{equation}
The values for the parameters are: $a=0.3$, $p=0.707$ and
$A=0.322$. The parameter $A$ is chosen such that the function $f(\nu)$
is normalised to unity. The \citet{press1974} mass function is recovered
when $a=1$, $p=0$ and $A=1/2$.

\subsection{Galaxy bias}

In the Lagrangian space the overdensity of halos $\delta_h(m,z)$ of mass
$m$ is related to the matter overdensity $\delta$ through the bias
relation. In the case of deterministic local linear bias the halo
overdensity is $\delta_h(m,z) = b(m,z) \delta$. It is natural to think
that galaxies are formed inside halos where the conditions for galaxy
formation exists. The relation between halos and galaxies is not
straightfoward due to the complexity of the galaxy formation
process. To deal with this problem the number of galaxies within a
halo $N_{g}$ is considered as a stochactic variable depending on the
mass of the halo. The distribution of $N_{g}$ is called the Halo
Occupation Distribution (HOD) \citep{seljak2000}. If in a catalogue
there are only halos of mass greater than $M_{\rmn{min}}$ then galaxy
overdensity $\delta_g(z)$ is given by
\begin{equation}
\label{eqn:halo_overdensity}
\delta_g(z) = \frac{\int_{M_{\rmn{min}}}^\infty \mathrm{d}m \ n(m,z) \langle
  N_{g} \rangle \delta_h(m,z) }{\int_{M_{\rmn{min}}}^\infty \mathrm{d}m
  \ n(m,z) \langle N_{g} \rangle} \ , 
\end{equation}
where $n(m,z)$ is the halo mass function (Gausian or non-Gaussian,
depending on the case considered).
The upper limit in the integrals is taken to be infinity considering that
there are halos of arbritary large mass in the sample. In practice the
results are not strongly dependent on this maximum mass. The
equation (\ref{eqn:halo_overdensity}) leads to a similar formula for
the galaxy bias:
\begin{equation}
\label{eqn:bias}
b_g(z) = \frac{\int_{M_{\rmn{min}}}^\infty \mathrm{d}m \ n(m,z) \langle N_{g} \rangle
  b(m,z) }{\int_{M_{\rmn{min}}}^\infty \mathrm{d}m \ n(m,z) \langle N_{g} \rangle} \ .
\end{equation}
Then we have that $\delta_g(z) = b_g(z) \delta$. The halo bias depends
on the mass and the redshift of the halo. For the Gaussian bias, as
well as for the mass function, the expression in \citet{sheth1999} is
adopted:
\begin{equation}
b(m,z) = 1 + \frac{1}{D(z)\sigma(m)} \left( a \nu - \frac{1}{\nu} +
\frac{2p/\nu}{1+(a\nu^2)^p} \right) \ ,
\end{equation}
where $D(z)$ is the linear growth factor of perturbations normalised
to unity at $z=0$. The equation (\ref{eqn:bias}) depends on the
average number of galaxies within a halo $\langle N_g \rangle$. As
expected this quantity depends on the mass of the halo. We assume a
power-law model \citep{berlind2002},
\begin{equation}
\label{eqn:ng}
\langle N_g \rangle = \left\{
\begin{array}{ll}
0            & \mbox{if $M < M_{\rmn{min}}$} \\
(M/M_0)^\beta & \mbox{if $M > M_{\rmn{min}}$}
\end{array} \right. \ .
\end{equation}
The parameter $\beta$ is positive definite in order to have an
increasing number of galaxies with mass. In this formula the pivot
of the power-law $M_0$ represents the mass of the halos with an
average number of galaxies equal to one. The equation
(\ref{eqn:bias}) does not depend on the parameter $M_0$ and then it
can not be constrained using the power spectrum data.
The minimum mass of halo with galaxies $M_{\rmn{min}}$ is the same as
in equation (\ref{eqn:bias}).

\subsection{Galaxy distribution and HOD}

It is possible to deduce the galaxy distribution from the mass
function of the halos. The mass function gives the number of halos
per unit volume and mass. Multiplying the mass function by the average
number of galaxies $\langle N_g \rangle$ we get the number of
galaxies per redshift interval within a solid angle $\Omega$:
\begin{equation}
\der{n}{z} = \Omega \frac{r(z)^2}{H(z)} \int_0^\infty \rmn{d}m
\ n(m,z) \langle N_g \rangle  \ ,
\end{equation}
where $r(z)$ is the comoving distance and $H(z)$ is the hubble
function. The factor before the integral is the volume per redshift
interval.

\begin{figure}
\centering
\includegraphics[width=8cm]{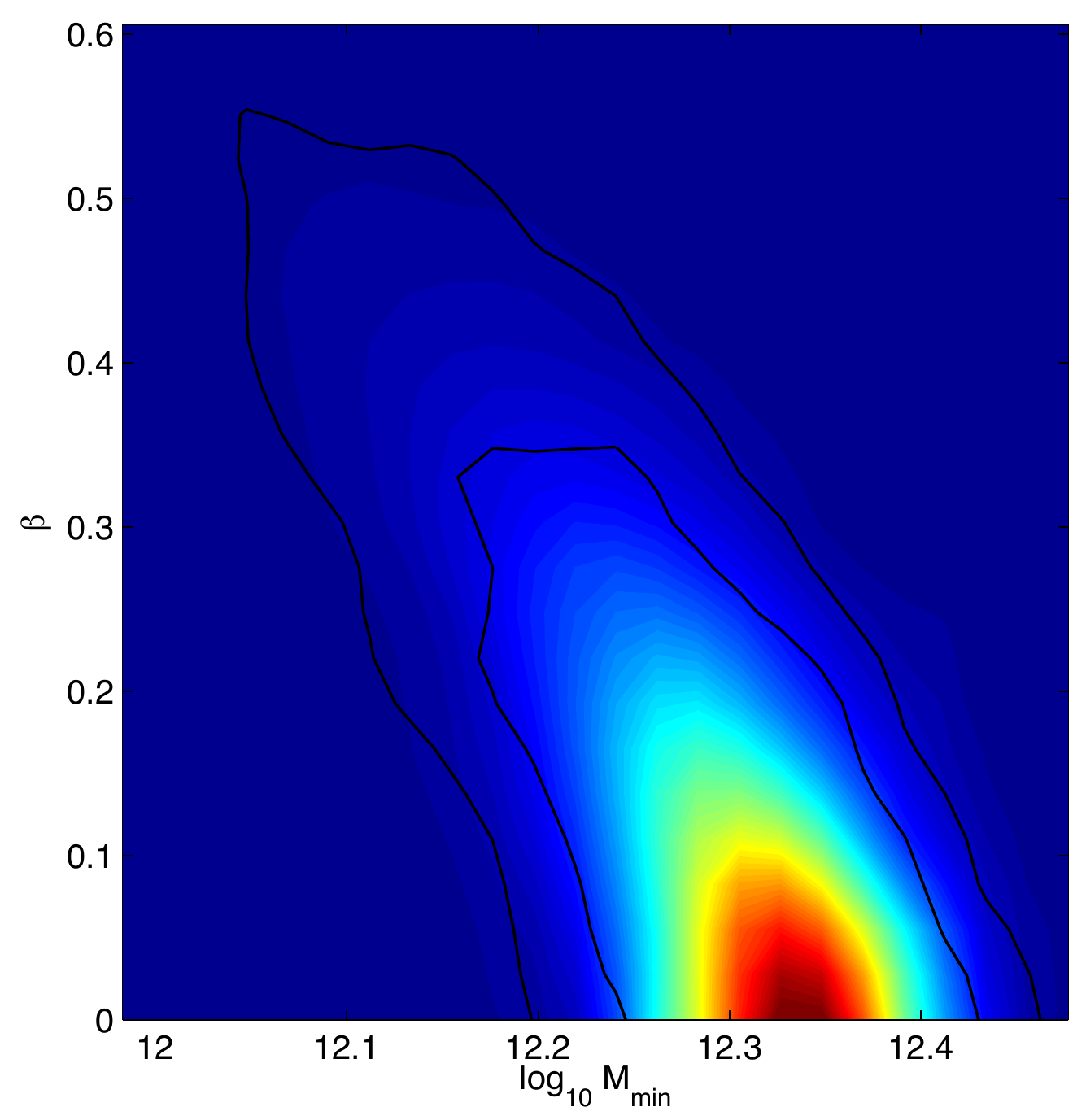}
\caption{Parameters concerning the HOD.}
\label{fig:hod}
\end{figure}

The theoretical power spectrum can be completely given by the
halo model and the HOD. It is possible to put constraints on the HOD
parameters using the power spectrum data of NVSS. Only the minimum
mass $M_{\rmn{min}}$ and the slope parameter $\beta$ can be
constrained using correlation functions. In Fig.~\ref{fig:hod} it is
represented the posterior probability of these two parameters. With
the NVSS data we only can conclude that there is not detection
of the slope and the upper limit is $\beta < 0.24$ ($1$-$\sigma$
c.l.). For this reason in the following we will assume the slope
paremeter is negligible.

\subsection{Galaxy angular power spectrum}

In order to study two-dimensinonal surveys like NVSS is necessary to
project the tridimensional galaxy power spectrum in the sphere. The
angular power spectrum is calculated from the linear power spectrum
once it is provided the redshift distribution. The galaxy angular
power spectrum is given by 
\begin{equation}
C_\ell = \frac{2}{\pi} \int \rmn{d}k \ k^2 I_\ell^2(k) \ P(k) \ ,
\end{equation}
where $P(k)$ is the matter power spectrum at $z=0$ and
\begin{equation}
I_\ell(k) = \int_0^\infty \rmn{d}z \ b_g(z) \frac{\rmn{d}n}{\rmn{d}z}
D(z) \ j_l(k\,r(z))
\end{equation}
is the filter fuction of galaxies. In this expression $j_\ell$ are the
spherical Bessel functions and $r(z)$ is the comoving distance as a
function of redshift. The redshift distribution is multiplied by the
galaxy bias $b_g$ and matter overdensity growth factor $D(z)$.

\subsection{Primordial non-Gaussianity}

Using the Large Scale Structure (LSS) of the universe it is possible to
constrain the primordial non-Gaussianity
\citep{dalal2008,desjacques2010}. The nature of the primordial
perturbations affects to the distribution of dark matter halos. In
particular non-Gaussianity in the matter distribution modifies the
two-point correlation function of halos. The bias $b$ depends on time
(or redshift), but in the case of initial Gaussian perturbations it is
independent of the scale $k$. However the correction to the bias due
to primordial non-Gaussianity is scale-dependent, and this is an
important feature of the non-Gaussian bias.

In the local non-Gaussianity the potential is given by
\begin{equation}
\label{eqn:local_ng}
\Phi(\mathbf{x}) =  \phi(\mathbf{x}) + f_{\rmn{NL}} \left(
\phi^2(\mathbf{x}) - \langle \phi^2 \rangle \right) \ ,
\end{equation}
where the scalar $\phi$ is a Gaussian random field and $f_{\rmn{NL}}$ is the
non-linear coupling parameter which multiplies a term cuadratic in
$\phi$. In the case that $f_{\rmn{NL}}$ vanishes the potential is
Gaussian.  Following scalar perturbation theory the matter
perturbations $\delta$ are related with the potential through the
Poisson equation ($\nabla^2 \Phi \propto \delta$). In Fourier space
this leads to
\begin{equation}
\delta_m(k,z) = \mathcal{M}_m(k,z) \Phi(k) \ .
\end{equation}
The perturbation field $\delta_m$ is the filtered matter perturbation
field $\delta$ at a scale corresponding to halos of mass $m$ (top-hat
filter is assumed). The factor $\mathcal{M}_m(k,z)$ is
\begin{equation}
\label{eqn:phi_delta_transfer}
\mathcal{M}_m(k,z) = \frac{2}{3} \frac{k^2 T(k) D(z)}{\Omega_m H_0^2}
\frac{1}{g(z=\infty)} W_m(k) \ ,
\end{equation}
where $T(k)$ is the matter transfer function and $D(z)$ is the linear
growth factor of matter normalised to unity at redshift zero. The
quantity $g(z=\infty)$ is the growth suppression rate at redshift
infinity. This number gives the decay of the potential in non-matter
dominated universes, $\Phi(z) = g(z) \Phi(z=0)$. It is related to the
linear growth factor as $g(z) = D(z)(1+z)$. The reason for including
this term is that we are assuming the CMB convention for the $f_{\rmn{NL}}$
parameter, that is, in the equation (\ref{eqn:local_ng}) all the
fields are evaluated at redshift infinity. In the literature it is
also used the LSS convention is which the fields are given at redshift
zero, and then this term does not appear. For the currently favoured
$\Lambda$CDM model $g(z=\infty)$ has a value approximately equal to
$1.4$. The function $W_m(k)$ in equation
(\ref{eqn:phi_delta_transfer}) is the Fourier transform of the top-hat
filter at scale of halos with Lagragian mass $m$.

The expression for the non-Gaussian bias $b_{NG}(m,z)$ is \citep{matarrese2008}
\begin{equation}
\label{eqn:ng_bias}
b_{NG}(m,z) = b(m,z) + 2 f_{NL} \left( b(m,z) - 1 \right)\delta_c(z)
\frac{\mathcal{F}_m(k)}{\mathcal{M}_m(k,z)} \ ,
\end{equation}
where the scale-dependent part is proportional to $f_{\rmn{NL}}$ and
it is given by the factor $\mathcal{M}_m(k,z)$ in equation
(\ref{eqn:phi_delta_transfer}) and the function $\mathcal{F}_m(k)$. In
the case of local non-Gaussianity it is
\begin{equation}
\begin{array}{rl}
\mathcal{F}_m(k) & = \frac{1}{8\pi^2\sigma_m^2} \int_0^\infty
\mathrm{d}k_1 \ k_1^2 \mathcal{M}_m(k_1) P_{\Phi}(k_1) \times \\
& \int_{-1}^{1} \mathrm{d}\mu \
\mathcal{M}_m(k_2) \left( \frac{P_{\Phi}(k_2)}{P_{\Phi}(k)}
+ 2 \right) \ ,
\end{array}
\end{equation}
where $k_2 = \sqrt{k^2 + k_1^2 + 2 k k_1 \mu}$ and $P_{\Phi}(k)$ is
the power spectrum of the potential which is related to the matter
power spectrum by $P_{\Phi}(k) = P_m(k,z) / \left[ \mathcal{M}_m(k,z)
\right]^2$.

The mass function of halos is also modified by primordial
non-Gaussianity. The correction to the Gaussian mass function is given
by a multiplicative factor \citep{matarrese2000,loverde2008}:
\begin{equation}
\label{eqn:ng_massf}
n_{NG}(m,z) = n_{G}(m,z) \ R(m,z) \ ,
\end{equation}
where
\begin{equation}
R(m,z) = 1 + \frac{\sigma_m}{6\nu} \left[ S_3 \left( \nu^4 - 2 \nu^2 -
  1 \right) - \der{S_3}{\ln \nu} \left( \nu^2 - 1 \right) \right] \ .
\end{equation}
In this expression $S_3$ is the skewness of the matter overdensity
field, and then it is proportional to $f_{\rmn{NL}}$.

\section{Data}

The main data used in this paper are the NVSS radio galaxies
catalogue. In order to calculate the theoretical angular power
spectrum it is necessary to know accurately the redshift distribution
of the NVSS counts. The redshift distribution is calculated from the
CENSORS data. Both are described in the next subsections.

\subsection{NVSS catalogue}

The NRAO VLA Sky Survey (NVSS) is a $1.4~\rmn{Ghz}$ survey of the
northern equatorial part of the sky up to $-40^\circ$ in declination
\citep{condon1998}. The most important contribution at this frequency
is provided by the Active Galactic Nuclei (AGN's). One important
feature of NVSS is the large sky coverage compared with other galaxy
surveys. This fact allows a better estimation of the angular power
spectrum.

The NVSS observations are made with two different array configurations
of the radio telescopes (D an Dnc). Depending on the declination
one of them is used. This fact introduces a declination
systematic in the NVSS catalogue. The mean density of counts depends
on the declination angle. In order to avoid this problem the NVSS sky
map is corrected following the next procedure: The map is divided in
declination bands with the same area. The total number of stripes in
the map is $70$. The mean number of counts is calculated in each
stripe. This number depends on the declination due to the
systematic. Then the pixels in the bands are rescaled such that the
mean number of counts in each band is equal to the full sky mean. That
is,
\begin{equation}
n^\prime_{ai} = \frac{\bar{n}}{\bar{n}_a} \ n_{ai} \ ,
\end{equation}
where $\bar{n}_{a}$ is the mean number of counts in the band $a$,
$n_{ai}$ and $n^\prime_{ai}$ are the number of counts in the pixel $i$
of the band $a$ before and after the correction respectively. The mean
number of counts of the full sky is represented by $\bar{n}$. Note
that this transformation does not change the mean of the map, and the
shot noise is not affected by this tranformation.

\begin{figure}
\includegraphics[scale=0.3]{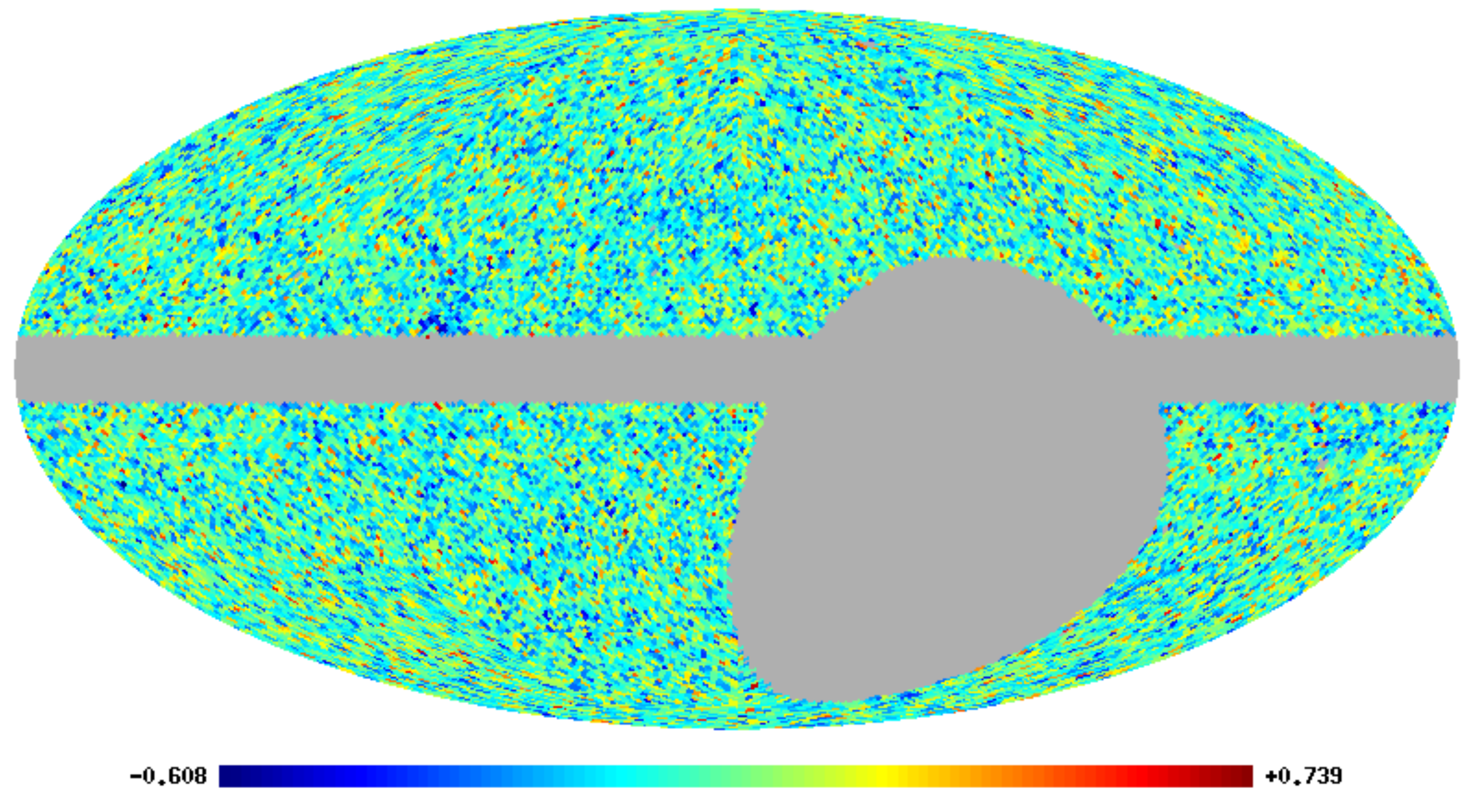}
\caption{
{\it Left}: NVSS map of sources with flux above $2.5~\rmn{mJy}$.
The map is at \healpix resolution $N_{\rmn{side}} = 64$ (pixel size
$\sim 1\degr$).
}
\label{fig:nvss_map_mask}
\end{figure}

The equatorial south pole from declination $-40\degr$ is masked due
to the absence of observations. In addition a region of width
$14\degr$ around the galactic plane is also masked to avoid
contamination from the Galaxy. Also some regions with high number of
counts are masked with a disk of $0.6\degr$ of radius. The resulting
mask is represented in Fig.~\ref{fig:nvss_map_mask}. The fraction of
observed sky is $f_{\rmn{sky}} = 0.73$.

We chose a flux threshold for the sources equal to
$2.5~\rmn{mJy}$. The total number of counts above this flux and
outside the mask is $1.45\e{6}$. It is important to have a high number
of counts in order to reduce the shot noise in the estimation of the
power spectrum.

\subsection{CENSORS catalogue}

For the modelling of the galaxy distribution of the NVSS catalogue it
is used the Combined EIS-NVSS Survey Of Radio Sources (CENSORS)
\citep{best2003,brookes2008}. This survey contains all the NVSS
sources above $7.2~\rmn{mJy}$ that are within a patch of $6~\deg^2$
in the ESO Imaging Survey (EIS). The total number of galaxies in the
survey are 149. The redshift of 44 sources are calculateed using $K-z$
relation, while the remainder 103 have spectroscopic redshift
\citep{brookes2008}. With these data it is possible to estimate the
redshift distribution of radio galaxies.

In the present analysis CENSORS data will be used to constrain the
redshift distribution of the NVSS radio galaxies. The accurate
knowledge of this distribution is importat for the calculation of the
angular power spectrum. The uncertainties in the estimation are taken
into account performing a joint fit of the galaxy distribution and
angular power spectrum at the same time. For this purpose the galaxy
distribution is modelled as:
\begin{equation}
\label{eqn:gamma}
\der{n}{z} = N \left( \frac{z}{z_0} \right)^\alpha e^{-\alpha z/z0}
\ .
\end{equation}
This gamma distribution depends on two parameters, $z_0$ and
$\alpha$. The physical meaning of $z_0$ is the redshift at which the
distribution is maximum, while $\alpha$ is a shape parameter. For
instance, the variance of the distribution is given by
$\frac{\alpha+1}{\alpha^2} z_0^2$. The constant $N$ is chosen such
that the distribution is normalised to unity. These two parameters
affect both the galaxy distribution and the angular power spectrum. In
Fig.~\ref{fig:galaxy_dist} it is represented the CENSORS data with a fit
using the gamma distribution. The best fit parammeters to the CENSORS
data are $z_0 = 0.53^{+0.11}_{-0.13}$ and $\alpha =
0.81^{+0.34}_{-0.32}$. In the literature there also exists a
parameterization of the CENSORS galaxy distribution given in
\citet{dezotti2010}:
\begin{equation}
\label{eqn:dezotti}
\der{n}{z} = 1.29 + 32.37 \, z - 32.89 \, z^2 + 11.13 \, z^3 - 1.25 \,
z^4 \ .
\end{equation}
It is a fit to the CENSORS data using a fourth order polynomial. The
units of this equation are number of counts per square degree.

\section{Data analysis}

In this work we use the angular power spectrum from NVSS and the
galaxy distribution given by CENSORS. In this section we describe the
procedure followed in the analysis of these two data sets. Also in
this section the likelihood function used to constrain the models is
shown.

\subsection{Angular power spectrum}

The estimation of the angular power spectrum has the problem of the
masked sky. The mask introduces correlations between multipoles and
also affect the estimate of the power spectrum. The estimator of the
angular power spectum is given by \citep{hivon2002}
\begin{equation}
\hat{C}_\ell = \sum_{\ellp m^\prime} M_{\ell\ellp} \frac{1}{2\ellp+1}
| a_{\ellp m^\prime} |^2 \ ,
\end{equation}
where the $a_{\ell m}$'s are the spherical harmonic coeficients of the
masked map. The mode coupling matrix $M_{\ell\ellp}$ is
\begin{equation}
\label{eqn:master}
M_{\ell_1\ell_2} = \frac{2 \ell_2 + 1}{4\pi} \sum_{\ell_3} w_{\ell_3}
\left( \begin{array}{ccc}
\ell_1 & \ell_2 & \ell_3 \\
 0 & 0 & 0
\end{array} \right)^2 \ .
\end{equation}
The coefficients $w_\ell$ are the multipoles of the angular power
spectrum of the mask. For the theoretical angular power spectrum
$C_\ell$ a Gaussian likelihood is supposed:
\begin{equation}
\label{eqn:cls_likelihood}
- \ln \mathcal{L} = \frac{1}{2} \sum_{\ell,\ellp}( \hat{C}_\ell - C_\ell )
\ F_{\ell\ellp} \ ( \hat{C}_{\ellp} - C_{\ellp}
) - \frac{1}{2} \ln |F| \ ,
\end{equation}
where the Fisher matrix $F_{\ell\ellp}$ is the inverse of the
covariance matrix of the $\hat{C}_\ell$'s. The theoretical power
spectrum $C_\ell$ as well as the Fisher matrix depend on the
parameters of the model. In the case of the full sky approximation
this matrix is diagonal, but in general the mask introduces couplings
between multipoles. In a theoretical framework the Fisher matrix is
given by \citep{hinshaw2003,xia2011}
\begin{equation}
F_{\ell\ellp} = \frac{(2 \l +1) M_{\ell\ellp}}{(C_\ell
  + N_\ell) (C_{\ellp} + N_{\ellp})} \ ,
\end{equation}
where the matrix $M_{\ell\ellp}$ is given by equation
(\ref{eqn:master}). We have verified that these estimates agree within
the uncertainties with those performed by the QML developed in
\citet{schiavon2012}.

\subsection{Galaxy distribution}

CENSORS data are divided in redshift bins each one of width 0.2. The
data cover a range of redshift from 0 to 3.6. In total there are 18
bins. The number counts in the $i$-th bin is denoted by
$\hat{n}_i$. We assume that the probability distribution of the number
counts in each bin is Poissonian:
\begin{equation}
\label{eqn:dndz_likelihood}
- \ln \mathcal{L} = \sum_i \left( n_i - \hat{n}_i \ln n_i  \right) \ ,
\end{equation}
where $n_i$ is the prediction of the model given by the galaxy
distribution in equation (\ref{eqn:gamma}). Note that we are not
taking into account correlations between different bins. Possible
correlations are negligible because the bin width is large compared
with the galaxy correlation scale and with the typical error in
redshift. Most of the galaxies have a spectroscopy error around
$10^{-3}$ \citep{brookes2008}.

\section{Results}

\subsection{Galaxy distributions}

\begin{figure}
\centering
\includegraphics[width=8cm]{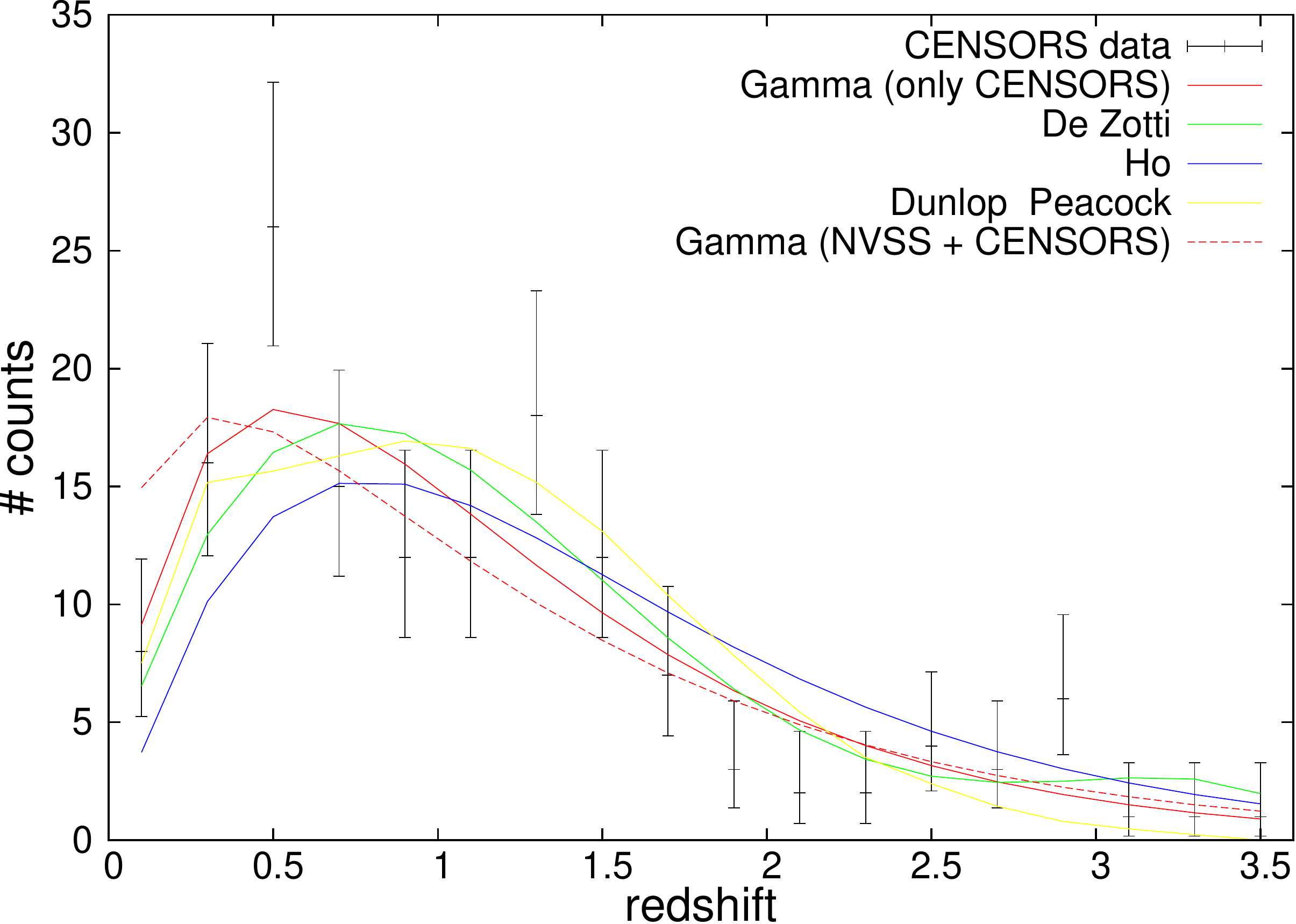}
\caption{Galaxy distribution for different models and CENSORS
  data. The error bars correpond to $68\%$ confidence level of the
  Poisson distribution. All distributions are nomalised such that the
  total number counts correspond to that of CENSORS. The red lines
  correspond to the gamma function model. The solid line is a fit to
  the CENSORS data and the dashed one also takes into account the
  NVSS power spectrum besides CENSORS.}
\label{fig:galaxy_dist}
\end{figure}

\begin{figure}
\centering
\includegraphics[width=8cm]{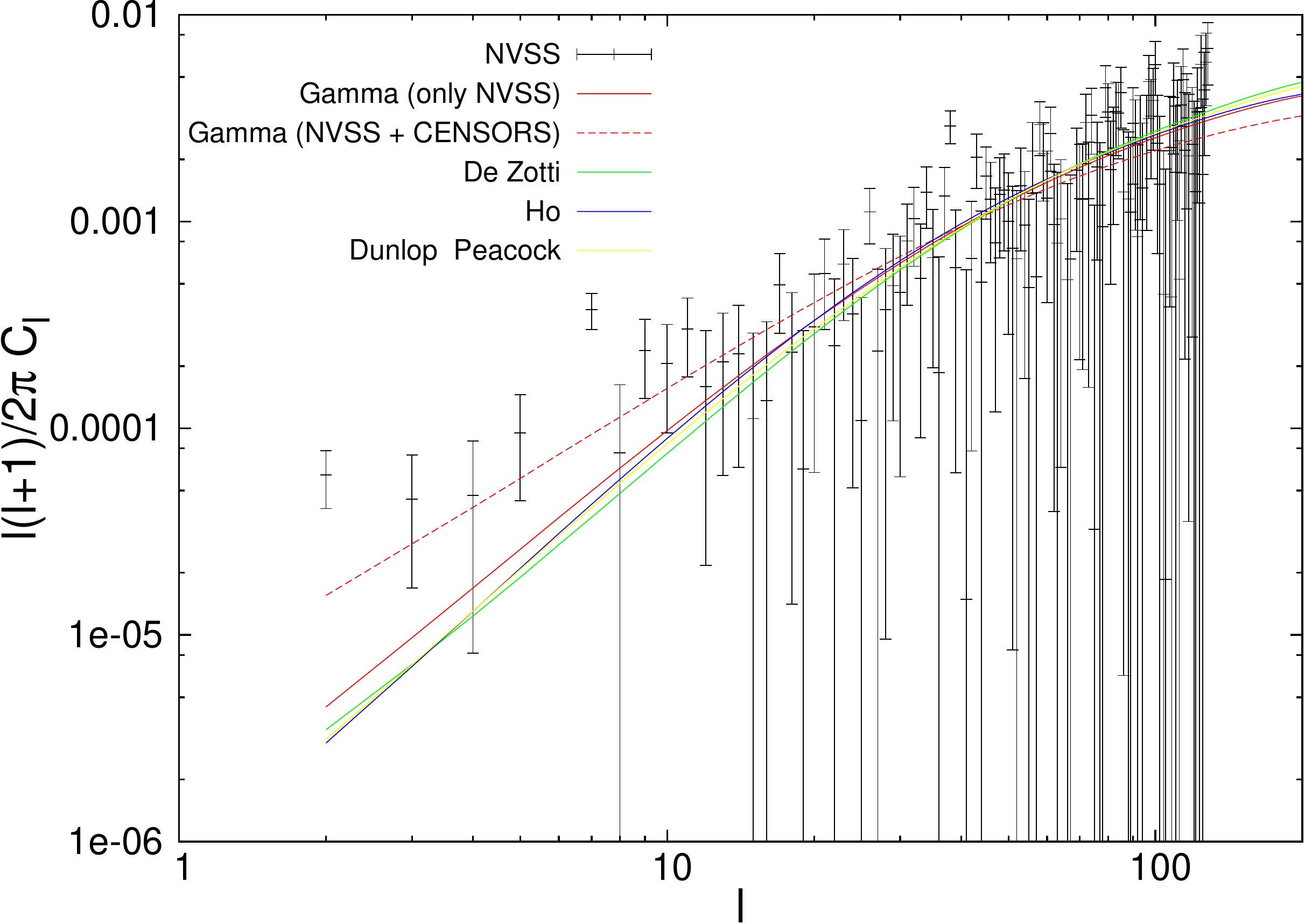}
\caption{Angular Power spectrum of NVSS. The points with the error
  bars represent the power spectrum of the NVSS sources with flux above
  $2.5~\rmn{mJy}$. The solid lines are different models proposed to
  describe the NVSS power spectrum ($\ell_{\rmn{min}} = 2$). The
  dashed line is a joint fit to the NVSS and CENSORS assuming a gamma
  function model for the galaxy distribution}
\label{fig:cls}
\end{figure}

In this section we will examine the galaxy distributions proposed to
describe the radio galaxies in the NVSS catalogue.  We use the NVSS
galaxy power spectrum as the data to compare the different galaxy
distribution models proposed in the literature. Different methods have
been used to model the NVSS sources. In \citet{dunlop1990} they
analyse the luminosity function of radio sources in order to determine
their evolution. For this model \citet{boughn2002} proposed a constant
bias model and they found the value of $b=1.6$. In \citet{ho2008} the
NVSS source distribution is calculate using the cross-correlation
between NVSS and other galaxy surveys, assuming a gamma function for
the fit. In the literature a parameterization of the CENSORS data from
\citet{dezotti2010} also exists. In this case the data are fitted to a
fourth order polynomial (see eqn. \ref{eqn:dezotti}). In
Fig.~\ref{fig:galaxy_dist} all these models for the NVSS galaxy
distribution are represented. We will perform a Bayesian evidence test
comparing all these models. The galaxy distribution is fixed and the
only free parameter is related to the bias. The model for the bias is
given by equation (\ref{eqn:bias}), except for the \citet{ho2008}
galaxy distribution. In this case it is assumed a constant bias as it
is suggested in that paper. The values of the evidences comparing all
the models are in the Table~\ref{tab:dndz_evidences}. It is examined
two cases: one when all the multipoles are taken into account
($\ell_{\rmn{min}} = 2$) and other considering only multipoles above
$\ell_{\rmn{min}} = 10$. When the full set of multipoles are
considered large differences between models are found, given more
evidence to the ``gamma model''. However if lower multipoles are
ignored the evidences are more similar. This fact could be an
indication of the inconsistency of the galaxy distribution and the
low-$\ell$ power spectrum data. In Fig.~\ref{fig:cls} it is
represented the power spectra for the different models. It can be
appreciated that there is a discrepancy between the models and data at
large scales.

\begin{table}
\centering
\begin{tabular}{@{}llcc}
\hline
& Model & $\ln(\rmn{Evidence})$ & $\ln(K)$ \\
\hline \multirow{4}{*}{$\ell_{\rmn{min}} = 2$}
& Gamma &               1472.68 & - \\
& \citet{dezotti2010} & 1459.15 & 13.53 \\
& \citet{ho2008} &      1465.28 & 7.4 \\
& \citet{dunlop1990} &  1463.62 & 9.06 \\
\hline \multirow{4}{*}{$\ell_{\rmn{min}} = 10$}
& Gamma &               1412.92 & - \\
& \citet{dezotti2010} & 1409.18 & 3.74 \\
& \citet{ho2008} &      1412.12 & 0.8 \\
& \citet{dunlop1990} &  1410.88 & 2.04 \\
\hline
\end{tabular}
\caption{Evidences for different models of the galaxy distribution
  taken into account the power spectrum data. The last column is the
  logarithm of the Bayes factor between the Gamma model and the
  others.}
\label{tab:dndz_evidences}
\end{table}

\subsection{Joint fit to NVSS and CENSORS data}

It is also possible to explore the case in which the NVSS power
spectum and the CENSORS galaxy distribution are taking into acccount
in the analysis at the same time. In this case the two data sets are
supposed to be independent and the joint likelihood is the product of
the likelihoods (\ref{eqn:cls_likelihood}) and
(\ref{eqn:dndz_likelihood}). For the galaxy distribution it is
supposed a gamma function model (equation \ref{eqn:gamma}). The
power spectrum also depends on the parameters affecting the galaxy
distribution as well as the bias parameters. As in the previous case we
assume a bias model depending on one parameter ($M_{\rmn{min}}$) given
by the equation (\ref{eqn:bias}). The best fit power spectrum is shown
in Fig.~\ref{fig:cls}.

In Table~\ref{tab:cls_dndz_evidences} it is shown the evidences of the
joint fit (varying also the galaxy distribution parameters) and models
with the galaxy distributions fixed. The same galaxy distributions
that appear in Table~\ref{tab:dndz_evidences} are considered. We can
see that the results derived for the two cases considered, all
multipoles and only multipoles $\ell \ge 10$, are quite different. The
evidences are more similar in the first case and this is another prove
that something strange is present at large scales. When all the
multipoles are considered the joint fit provides much higher evidence
than the other models with less number of parameters. This is a
consecuence of the freedom of the joint fit model to reproduce the
problematic data. The parameters for this case appear in the first
rows of Table~\ref{tab:ng_comparison} (Gaussian case). The redshift of
the maximum of the distribution ($z_0 = 0.33$) differs from the one
obtained with the CENSORS data only ($z_0=0.53$) and the error bars
are not compatible. This fact is an evidence of some tension between
the NVSS and the CENSORS data sets, possibly caused by the excess of
power in the NVSS power spectum. The increment of the number counts at
low redshifts produces more power at large scales in the angular power
spectrum.

\begin{table}
\centering
\begin{tabular}{@{}llcc}
\hline
& Model & $\ln(\rmn{Evidence})$ & $\ln(K)$ \\
\hline \multirow{5}{*}{$\ell_{\rmn{min}} = 2$}
& Joint fit &           1436.91 & - \\
& Gamma &               1430.63 & 6.28 \\
& \citet{dezotti2010} & 1415.77 & 21.14 \\
& \citet{ho2008} &      1415.50 & 21.41 \\
& \citet{dunlop1990} &  1412.05 & 24.86 \\
\hline \multirow{5}{*}{$\ell_{\rmn{min}} = 10$}
& Joint fit &           1365.37 & - \\
& Gamma &               1370.86 & -5.49 \\
& \citet{dezotti2010} & 1365.79 & -0.42 \\
& \citet{ho2008} &      1362.35 &  3.02 \\
& \citet{dunlop1990} &  1359.31 &  6.06 \\
\hline
\end{tabular}
\caption{Evidence for the comparison of different galaxy distribution
  models with the joint fit model using the power spectrum and
  CENSORS data. The last column is the logarithm of the bayes factor
  between each galaxy distribution and the joint fit.}
\label{tab:cls_dndz_evidences}
\end{table}

\subsection{Non-Gaussianity}

We explore non-Gaussianity in NVSS taken into account also the CENSORS
data. A joint fit of the NVSS power spectrum and the galaxy
distribution provided by CENSORS is done in a scenario with primordial
non-Gaussianity. The primordial non-Gaussianity is more significant at
large scales and for this reason all the multipoles are included in
this analysis. As mentioned before, primordial non-Gaussianity is a
possible explanation for the excess of power at large scales.

\begin{figure}
\centering
\includegraphics[scale=0.48]{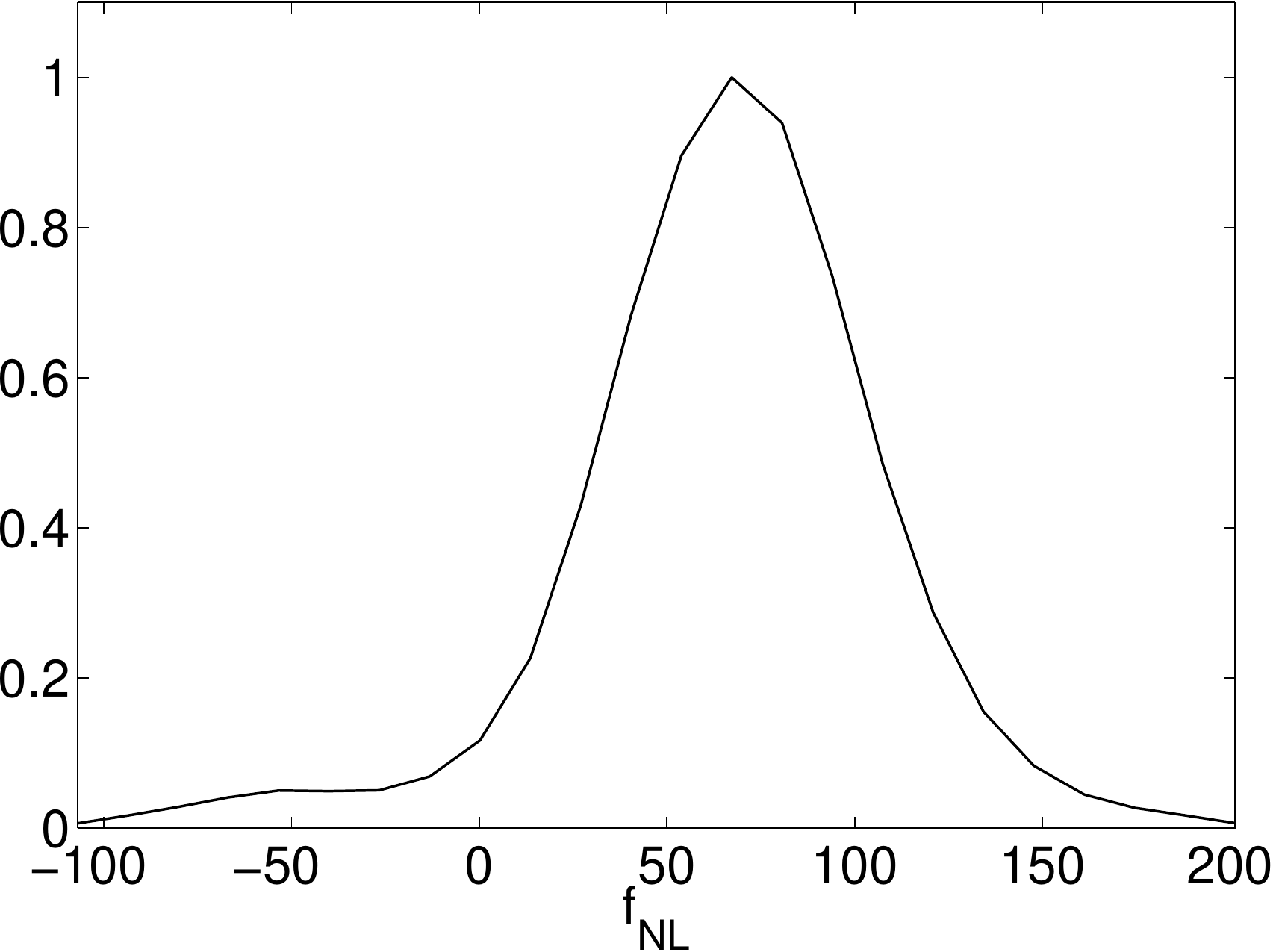}
\caption{Marginalised likelihood for $f_{\rmn{NL}}$.}
\label{fig:like_fnl}
\end{figure}

In the Table~\ref{tab:ng_comparison} the parameters obtained with and
without non-Gaussianity are summarized. In the non-Gaussian case a
$f_{\rmn{NL}}$ of approximately $54$ is needed to fit the
data. However the $2$-$\sigma$ constraints, $-43 < f_{\rmn{NL}} <
142$, are compatlble with the Gaussian case ($f_{\rmn{NL}} = 0$) and
also with the result obtained in \citet{planck24_2013}
($f_{\rmn{NL}}=2.7 \pm 5.8$). The value of $f_{\rmn{NL}}$ that we find
is more consistent with zero than the one obtained in \citet{xia2010}
($25<f_{\rmn{NL}}<117$) and it is compatible with the result in
\citet{giannantonio2013} for the NVSS power spectrum. In
Fig.~\ref{fig:like_fnl} it is represented the marginalised probability
of $f_{\rmn{NL}}$. Also in the last column of the
Table~\ref{tab:ng_comparison} there are the evidences of both
models. These numbers provide a Bayes factor $\ln(K)=1.25$ and
therefore there is no strong preference for any model. But when
non-Gaussianity is taken into account the parameters of the galaxy
distribution, specially $z_0$ (the maximum of the distribution), are
compatible with the CENSORS data. Then, the tension between the NVSS
and the CENSORS data sets is reduced.

\begin{table}
\begin{tabular}{@{}llcccc}
\hline
& Parameter & best fit & Mean  & $\sigma$ & $\ln(\rmn{Evidence})$ \\
\hline \multirow{3}{*}{G}
& $\log_{10} M_{\rmn{min}}$  & 12.66 & 12.63 & 0.19 &
\multirow{3}{*}{$1437.59 \pm 0.13$} \\
& $z_0$                  & 0.33  & 0.34 &  0.04 & \\
& $\alpha$               & 0.37  & 0.40 &  0.09 & \\
\hline \multirow{4}{*}{NG}
& $\log_{10} M_{\rmn{min}}$ & 12.52 & 12.44 & 0.20  &
\multirow{3}{*}{$1438.84 \pm 0.13$} \\
& $z_0$                 &  0.46 &  0.49  &  0.08 & \\
& $\alpha$              &  0.63 &  0.75  & 0.22 & \\
& $f_{\rmn{NL}}$               &  54   &  66    &  40  & \\  
\hline
\end{tabular}
\caption{Comparison of the two models with and without
  non-Gaussianity. Best fit  and the marginalised mean vaule with the
  1-$\sigma$ errors. The last column is  the evidence of each
  model. These values are for the full set of multipoles
  ($\ell  \ge 2$)}
\label{tab:ng_comparison}
\end{table}

\section{Conclusions}

In this paper we have analysed the NVSS power spectrum trying to
understand the discrepancy existing between data and the theoretical
model at large scales. There are some problems concerning the NVSS
data. The first of them is a declination systematic present in the
number counts of NVSS due to the observation strategy of the VLA. When
fainter sources are removed from the catalogue this systematic becomes
less significant. We consider that the declination systematic is
negligible for sources with flux above $10~\rmn{mJy}$, but it remains
for data with lower flux threshold. In order to mitigate this problem
we propose a correction of the data based on the rescaling of the
number of counts in declitation stripes. This correction is applied to
NVSS counts with flux above $2.5~\rmn{mJy}$. When we compare the power
spectra for different flux thresholds they are compatible. The advance
of working with sources with a flux limit of $2.5~\rmn{mJy}$ is that
the Poissonian noise is considerably reduced. Having under control the
declination systematic the problem of NVSS at large scales still
persists. In this work we consider different theoretical models for
the NVSS galaxy distribution and all of them have problems in the
power spectrum at large scales. When the large scales are removed from
the analysis ($\ell \ge 10$) then all the models become compatible
with the data. This fact is a direct evidence of the lack of a good
modelling of NVSS at large scales.

There are differrent possible explanations for exccess of power at
large scales. It could be an unexplained systematic in the NVSS
data. In the present work we rule out the declination systematic as
the origin of this excess. Another possibility is that the origin of
the excess is fundamental. A well known mechanism which contributes to
the power spectrum at large scales is the primordial
non-Gaussianity. We have also tested the possibility that primordial
non-Gaussianity could explain the power excess. The value infered for
$f_{\rmn{NL}}$ has a large error bar and it is compatible with zero at the
$2$-$\sigma$ confidence level. For values of $f_{\rmn{NL}}$ compatible with
the Planck constraints \citep{planck24_2013} the effect on the fit of
the data is negligible.

\section*{Acknowledgments}

We acknowledge partial financial support from the Spanish \textit{Ministerio de Econom\'ia y Competitividad} Projects AYA2010-21766-C03-01, AYA2012-39475-C02-01 and Consolider-Ingenio 2010 CSD2010-00064. The authors acknowledge the computer resources, technical expertise and assistance provided by the \textit{Spanish Supercomputing Network} (RES) node at Universidad de Cantabria.

\bibliographystyle{mn2e}
\bibliography{nvss}


\bsp

\label{lastpage}

\end{document}